
\hsize=5 true in
\vsize=7.8 true in
\baselineskip 13 pt
\font\eightrm=xcmr8
\font\eightit=xcmti8
\font\ninerm=xcmr9
\font\nineit=xcmti9
\font\ninebf=xcmbx9
\font\boldit=xcmbxti1

\def\oh {{1\over2}}
\def\ni {\noindent}
\def\a {\alpha}
\def\b {\beta}
\def\g {\gamma}
\def\G {\Gamma}
\def\del {\delta}
\def\e {\epsilon}
\def\f {\varphi}
\def\la {\lambda}

\def\om {\omega}
\def\Om {\Omega}
\def\d {\partial}
\font\simv=cmsy10
\def\M{\hbox{\simv \char 77}}
\def\t {\tilde}
\def\na {\nabla}

\def\GG #1 #2 #3 {\Gamma ^{\ #2}_{#1 \ #3}}
\def\R #1 #2 #3 #4 {R^{#1}_{\ #2 #3 #4}}
\def\E #1 #2 {E^{#1}_{\ #2}}
\def\F #1 #2 #3 #4{F^{r}_{\ s \a \b}}

\def\i {\item}

{\baselineskip=10 pt
 \eightrm
 \leftline{Modern Physics Letters A, Vol. 0, No. 0 (1994) 000--000}
 \leftline{ \copyright , World Scientific Publishing Company}}

\vskip 48 pt
\centerline{\bf FOUR DIMENSIONAL YANG - MILLS THEORY}
\centerline{\bf IN LOCAL GAUGE INVARIANT VARIABLES}

\vskip 24 pt

\centerline{F.A.LUNEV \footnote{${}^{\#}$}{\eightrm   e-mail
address:  lunev@hep.phys.msu.su} }

\vskip 12 pt

\centerline{\eightit Physical Department, Moscow State
University,}

\centerline{\eightit Moscow,119899,Russia}

\vskip 20 pt

\centerline{\eightrm Received }

\vskip 16 pt

{\rightskip=0.8 cm
 \leftskip=0.8 cm
{ \eightrm
 \baselineskip=10 pt

It is  presented the  general method that allows to formulate
4D $SU(N)$ Yang - Mills theory in terms of only local gauge
invariant variables. For the case N=2, that is discussed in
details, this gauge invariant formulation appears to be very
similar to $ R^2 $-gravity.}}

\vskip 24 pt

\noindent {\bf 1.Introduction.}

\noindent In previous author's works it was given formulation of three
dimensional $SU(2)$ Yang - Mills (YM) theory in terms of local
gauge invariant variables both at classical [1], [2] and quantum
levels [15] \footnote{$ ^a $}{\eightrm \baselineskip=10 pt It is
necessary to note that similar construction for selfdual sector
of 4D Euclidean YM theory was given in works [21]} .  It was
found that obtained formulation is very similar to 3D Einstein
gravity.  Using analogy with gravity, it was discovered new
static solution of classical four dimensional YM equations with
singularity on the finite sphere that is analog of Schwarzschild
solution in general relativity and represents confinement
potential for quantum particle moving inside this sphere.

  In this paper we  will give generalization of these results for
4D case and $SU(N)$ gauge group for arbitrary $N$. As in 3D case
we will find that in proposed gauge invariant variables YM
theory seems very similar to gravity but, in contrast to 3D
case, to $R^2$-gravity rather than Einstein one. This analogy
with gravity was already exploited to obtained new interesting
solutions of coupled Einstein - Yang - Mills - Higgs equations
[16].

Gauge invariant formulation of three dimensional $SU(2)$ YM
theory presented in[1] is, in fact, nothing else but realization
of Lie - Vessiot \footnote {$ ^b $}{\eightrm \baselineskip=10 pt
Modern formulation of Lie - Vessiot theory see [3]}
decomposition of YM equations with respect to gauge group. In
principle, using Lie - Vessiot method, one can reformulate in
invariant variables any system of differential equations that is
invariant under the action of certain group of point
transformation. But resulting formulation is not, in general,
Lagrangian and manifestly relativistic covariant. In particular,
in contrast to 3D case, direct application of Lie - Vessiot
method to 4D YM equations leads to the theory formulated in
terms of gauge invariant but relativistic noncovariant
quantities such as $ F^a_{oi} F^a_{oj}, \  F^a_{ij} F^a_{kl}$,
etc.. \footnote{$ ^c $}{\eightrm \baselineskip=10 pt Formulation
of $SU(2)$ YM theory in terms of the variables $ F^a_{ij}
F^a_{kl}$ in gauge $A_0=0$ was given recently in papers [17] }

In order to overcome these difficulties, we will modify the
methods of [1] in the following way. First, we will show that
initial $SU(N)$  YM theory is equivalent to certain $GL(N,C)$
gauge theory in the following sense: classes of gauge equivalent
solutions of $SU(N)$ YM theory are in one-to-one correspondence
with classes of gauge equivalent solutions of above mentioned
$GL(N,C)$ gauge theory. Further, using the fact, that Lorentz
group $SL(2,C)$ is isomorfic to subgroup of gauge group
$GL(N,C)$, it is already easy to reformulate this $GL(N,C)$
gauge theory in gauge invariant and relativistic covariant
variables.

The paper is organize as follows. In the section 2 we discuss
some auxiliary questions concerning gauges of unitary type. In
section 3 we present our main results about gauge invariant
formulation of $SU(N)$ YM theory. In the section 4 we consider in
details the case $N=2$. In the section 5 we briefly discuss the
case $N \geq 2 $, gauge invariant formulation of YM theory in
Eucleadian space, introducing of matter fields, and 3D case. In
the last section we discuss the themes of further investigations.

\vskip 23 pt

\noindent {\bf 2. Ultralocal fields and generalized unitary gauges.}

\noindent Let us consider field theory with Lagrangian $ L=L[ \f
], \ \f=(\f ^m; \ m=1,...,M) $ that is invariant under $ R $
parametric gauge group $ G $ with group parameters $ \om = ( \om
^r(x), r=1,...,R) $. Let $ \f ^{\om} = ((\f ^{\om})^m ) $ is
gauge transformation of the field $ \f $ . In general,

$$ (\f ^{\om})^m = f^m (\f, \om, \d _{\a} \om, \d _{\a} \d _{\b}
\om,...) \eqno(1) $$

The field $\f^m$ is called ultralocal (UL), if

$$ {\d f^m \over \d (\d _{\a} \om ^r ) } = 0, \ {\d f^m \over
\d (\d _{\a} \d _{\b} \om ^r ) } =0, \ ... \eqno(2)$$

For example, field strength tensor $F^r_{\a \b } $ is ultralocal
field, but potentials $ A^r_{\a } $ are not ultralocal.

Under infinitesimal gauge transformations UL fields transforms
as follows:

$$ \del \f ^m = R^m_{\ r} (\f) \del \om ^r \eqno(3)$$

By definition, generalized unitary (GU) gauge is the
condition

$$\f ^a =0, \ \ a=1,...,R_0 \leq R \eqno(4) $$

\ni where $\f ^a $ - ultralocal fields. For example, usual
unitary gauge

$$ \f ^2 =0, \ Im \ \f ^1 =0 \eqno(5) $$

\ni in Salam - Weinberg model [4] with doublet of Higgs bosons
$\f ^1, \f ^2 $ is GU gauge, but it is not so for Lorenz gauge $
\d _{\a} A^{r \a} =0 $.

Let $ \M $ be a family of field configurations that is invariant
under the action of the gauge group $ G $. The gauge (4) is
called regular on $ \M $, if

$$ \hbox { rank } (R^a_{\ r})_{\left | \matrix {\f ^a =0 \cr \f
^m \in \M } \right. } =R_0 \eqno(6)$$

\ni for almost all $ x $. For example, usual unitary gauge (5)
is regular on any gauge invariant set $\M$ that doesn't contain
zero configuration $\f ^a \equiv 0 $ of Higgs fields.

Further, let

$$ \t L = \t L[\f] = L[\f]_{\left| \matrix { \f ^a =0
 } \right. }  \eqno (7) $$

\ni The following proposition will be often used below:

{\it   {Let $ \ \M$ be the set of solutions of equations }}

$$ {\del \t L \over \del \f ^m } =0, \ m=R_0 + 1,...,M \eqno(8) $$

{\it {for which the condition (6) is hold. Then any field $
\f \in \M $ satisfy also the equations}}

$$ \left. {\del L \over \del \f ^m } \right. _{ \left | \f ^a =0
\right. } =0, \ \ a=1,...,R_0, \ m=1,...,M \eqno(9)$$

The proof is very simple. Indeed, for $ m=R_0 +1,...,M $

$$ \left. {\del L \over \del \f ^m } \right. _{ \left | \f ^a =0
\right. } = {\del \t L \over \del \f ^m } =0 \eqno(10) $$

\ni So (9) is valid for $ m=R_0 +1,...,M.$ Further, due to the
second Noether theorem

$$ \sum _{a=1}^{R_0} R^a_{\ r} {\del  L \over \del \f ^a } +
\sum ^M_{m=R_0 + 1} {\hat R}^m_{\ r} {\del  L \over \del \f ^m }
= 0 \eqno(11) $$

\ni where ${\hat R}^m_{\ r}$ are some linear differential
operators. So from (6), (10) and (11) one can conclude that (9)
is valid also for $ a=1,...,R_0 $

Let $G_0 \subset G $ be the group of invariance of (4). If GU
gauge (4) is regular on $ \M$, then $G_0$ is transformation
group of $\M$ and $ \hbox {dim} G_0 = R - R_0 $ . In particular,
if $R=R_0$ then $G_0$ is discrete and so $G_0$ is the group of
global transformations.

Consequently, the proposition proved above means that classes of
$G_0$ equivalent solutions of equations (8) are in one-to-one
correspondence with classes of $G$-equivalent solutions of
initial equations

$$ {\del  L \over \del \f ^m }=0 \eqno(12)$$

\ni except, may be, the solutions of (8) for which (6) is not
hold.  But in any case the set of solutions of  equations (8) is
larger than the  set of solutions of (12). So one can use
Lagrangian $\t L$ instead of Lagrangian $L$ for investigation of
gauge theory with field equations (12).

If $R=R_0$ , then Lagrangian $\t L$ describes non gauge theory
because tne group $G_0$ is global. In fact, formulation of
the field theory by means of Lagrangian $\t L$ can be considered
as formulation in terms of gauge invariant variables \footnote
{ $ ^d $ } {\eightrm \baselineskip=10 pt Detailed discussion of
this statement from slightly different point of view can be
found in [5].}. Indeed, let us consider the change of variables

$$ (\f^1, ...,\f^M) \rightarrow (\om^1,...,\om^R, {\t
\f}^{R+1},...,{\t \f}^M) \eqno(13) $$

\ni where $ \om^a, {\t \f}^m $ are defined by equations

$$ (\f^{\om})^a = 0, \ \ a=1,...,R  \eqno(14) $$

$$ {\t \f}^m = (\f^{\om})^m, \ \ m=R+1,...,M    \eqno(15) $$

\ni (If (6) is hold, then (14) has only discrete set of solutions
and we can choose one of them).

Further, one can easy to prove, in new variables our theory is
described by Lagrangian

$$ \bar L [\om, \t \f] = L [\f] \eqno(16) $$

\ni Due to gauge invariance of $L$ and (14),

$$ L[\f] =L[\f ^{\om}] = L[\f]_{\left | {\f ^a = 0, \ \
a=1,...,R \atop \f ^m = {\t \f}^m , \ \ a=R+1,...M } \right. }
\eqno(17) $$

\ni So

$$ \bar L= \bar L [\t \f] = \t L [\t \f] \eqno(18) $$

Finally, it follows from (15), that ${\t \f}^m $ are gauge
invariant quantities. So the Lagrangian $\t L$ can be used for
description of given gauge theories in terms of gauge invariant
variables.

For example, for Salam - Weinberg model in gauge (5) $G_0 = Z_2$
and variables $\t \f, \t A_{\a} $ are defined up to $Z_2$
transformations by equations

$$ {\t \f}^2 = {\bar \f}^a \f ^a, \ \ \, \t A_{\a} = \Om A_
{\a} \Om ^{-1} + \Om \d _{\a} \Om ^{-1} \eqno (19) $$

\ni where bar denotes complex conjugation and

$$ \Om = ({\bar \f}^a \f ^a )^{- \oh} \left( \matrix {{\bar
\f}^1, & {\bar \f}^2 \cr - \f ^2 & \f ^1 } \right)$$

\ni Residue gauge freedom is fixed by boundary condition

$$ \t \f \to + \sqrt {{\bar \f}^a \f ^a}(\infty )= const \not =
0, \ \ |\vec x| \to \infty \eqno(20) $$

\ni Further discussion of Salam - Weinberg and more general
models from similar point of view see [5] or more recent works
[18], [19].

So, in order to formulate gauge theory in terms of local gauge
invariant variables, it is sufficient to impose generalized
unitary gauge. But such gauges don't exist in arbitrary gauge
theory. In particular, there are no GU gauges in pure YM theory
in standard second order formulation because of the absence of
ultalocal fields. In first order formalism, when field strength
tensor $ F_{\a \b} $ and potentials $ A_{\a} $ are considered as
independent variables, GU gauges exist. For instance, in recent
paper [6] for 4D  $ SU(2) $ YM theory it was proposed the GU
gauge

$$ F^a_{0i} = F^i_{0a} \eqno(21) $$

\ni Another, more complecated gauge of the type

$$ \Phi (F_{i0}) = 0 $$

\ni was given in [21].

But these gauges are not relativistic covariant. Moreover,
there is no relativistic covariant GU gauge in any YM theory
with compact gauge group without scalar fields (for example, for
QCD), because any GU gauge in such theories entangles space -
time and colour indexes (as in (21)) and it is impossible to
compensate transformation of noncompact Lorenz group by
transformations of compact gauge group in order to restore gauge
condition after Lorentz transformations.

To overcome these difficulties, in the next section we will
develop  formalism in which $SU(N)$ YM theory is change by
certain $GL(N,C)$ gauge theory that is equivalent to initial YM
theory in the sense explained in the section 1.

The group $GL(N,C)$ contains the Lorentz group $SL(2,C)$ as
a subgroup and this fact allows to use covariant GU gauge of the
type (21).

\vskip 23 pt

\noindent {\bf 3. The main construction: from $SU(N)$ YM theory to
$GL(N,C)$ gauge theory. }

\noindent Let vectors $ E_A= (E^r_{\ A}(x), \ A,r=1,...,N)$ for each
$x$
form the basis of the space of fundumental representation of Lie
algebra $su(N)$.  Let constant matrices

$$\la ^r = \left( {1 \over 2i}
(\la ^r)^s_{\ t} \right),\ \  \hbox {Tr} \la ^r \la ^s = - \del
^{rs} $$

\ni form the basis Lie algebra $su(N)$ in fundamental
representation,

$$ (\na _{\a})^r_{\ s} = \d _{\a} \del ^r_{\ s} + A^{\ r}_{\a \
s}, \ \ \ A^{\ r}_{\a \ s} =( A^t_{\a}\la ^t)^r_{\ s} \eqno(22)$$

\ni is usual covariant derivative and

$$F_{\a \b}= ([\na _{\a}, \na _{\b}]^r_{\ s})=(F^t_{\a \b}
\la ^t)^r_{\ s} \eqno(23) $$

\ni field strength tensor.

Now for given vectors $E^r_A$ we define quantities $\G^{\
A}_{\a \ B}, \ \ h_{AA'}$, and $\rho $ by equations

$$\na _{\a} E^r_{\ A} \equiv \d _{\a} E^r_{\ a}  + A^{\ r}_{\a \
s} E^s_{\ A} = \G^{\ B}_{\a \ A} E^r_{\ B} \eqno(24) $$

$$  h_{AA'}  =  E^r_{\  A}  {  \bar  E  }  ^r_{\ A'}, \ \ \rho =
 \hbox {det}
 (E^r_{\ A}) \eqno(25) $$

\ni  From definition

$$\d _{\a} h_{AA'} - \G^{\ B}_{\a \ A}h_{BA'} - {\bar \G} ^{\
B'}_{\a \ A'}h_{AB'} \equiv D_{\a}h_{AA'} =0 \eqno(26) $$

$$\d _{\a} \rho = \G^{\ A}_{\a \ A} \rho \eqno(27) $$

$$|\rho|^2 = \hbox {det} (h_{AA'}) \eqno(28) $$

Under the transformations $E^r_{\ a} \to E^r_{\ B} S^B_{\ A}, \
\  S^B_{\ A} \in GL(N,C)$, and $ \G^{\ B}_{\a \ A},  h_{AA'} $
are transformed as a connection and second rank tensor on
trivial fibre bundle $M^4 \times C^N$ where $M^4$ is space -
time manifold. It follows from (25),(26) that $h_{AA'}$ can be
considered as Hermitian positive defined metric on this bundle.

By definition, quantities $\GG {\a} A B , h_{AA'} $ and $\rho$
are $SU(N)$ are gauge invariant functions on $M^4$. Our next task
is reformulation of initial $SU(N)$ YM theory in terms of these
functions.

Let

$$ \R A B {\a} {\b} = [D_{\a}, D_{\b}]^A_{\ B} = \d _{\a}
\GG {\b} A B - \d _{\b} \GG{\a} A B + \GG{\a} A C \GG{\b} C B -
\GG{\b} A C \GG{\a} C B \eqno(29) $$

\ni be curvature tensor of connection  $\GG{\a} A B $. It is easy
to prove that

$$ \F r s {\a} {\b} \E s A = [\na_{\a}, \na_{b}]^r_{\ s} \E s A
= \R B A {\a} {\b} \E r B \eqno(30) $$

Using (30), one can express usual YM Lagrangian

$$ L_{YM} =- {1 \over 4e^2} \sqrt {-g} F^r_{\a \b} F^{r \a
\b} \eqno(31)$$

\ni through functions $\GG {\a} A B $:

$$ L_{YM} \equiv L[ \Gamma] = {1 \over 2e^2} \sqrt {-g}  \R A B
{\a} {\b} R^{B \ \a \b}_{\ A}          \eqno(32) $$

\ni where $g_{\a \b} $ is space - time metric tensor with
signature $(+---)$.)

Variables $ \GG {\a} A B $ are not independent because they
satisfy identities (26)-(28). Solving these identities with
 respect to $ \GG {\a} A B $ and substituting the result in (32),
 one obtains Lagrangian  of some $ GL(N,C) $ gauge theory.
 This Lagrangian depends on $ h_{AA'}, \rho $ and those of
functions $ \GG {\a} A B $ that can not be expressed trough $
h_{AA'}, \rho $ from identities (26)-(28).

Let $\M$ be the set of solutions of $GL(N,C)$ gauge theory with
Lagrangian (32) for which matrix  $h_{AA'}$ is  positive defined.
The set $\M$ is invariant under $GL(N,C)$ gauge transformations.

Now we can prove the following proposition, playing the central
role in our investigation:

{\it Classes of $GL(N,C)$ equivalent solutions from the set $\M$
of gauge theory with Lagrangian (32) are in one-to-one
correspondence  with classes of gauge equivalent solutions of
initial $SU(N)$ YM theory.}

Indeed, consider  $GL(N,C)$ gauge

$$ h_{AA'} = \del _{AA'}, \ \ \ \rho =1  \eqno(33) $$

It is   GU gauge that is regular on $\M$. It follows from
(26),(28) that matrices $\G _{\a} = ( \GG {\a} A B )$ satisfy
equations:

$$ \G _{\a} + \G ^{+}_{\a} =0, \ \ \  \hbox {Sp} \ \G _{\a} =0
\eqno(34)$$

But conditions (34) are exactly the definition of matrix Lie
algebra $su(N)$. So after substitution of (33) and (34) in (32)
we obtain usual YM Lagrangian (this operation is correct due to
proposition from section 2). Therefore in gauge (33) the set
$\M$ can be identifies with the set of all solutions of $SU(N)$
YM theory.

Further, solving (24) with respect to $ A^{\ r}_{\a \ s}$, one
obtains:

$$ A^{\ r}_{\a \ s} = \t E ^B_{\ s} \GG {\a} A B \E r A - \t E
^A_{\ s} \d _{\a} \E r s  \eqno(35) $$

\ni where $ \t E ^A_{\ r} \E r B = \del ^A_B $. But it follows
from (25) and (33) that $ \E r A \in SU(N) $. So potentials $
A_{\a} $ and $ \G _{\a} $ in gauge (33) are connected by $ SU(N)
$ gauge transformation. Consequently, if  $ \G _{\a} $ run over
the set of all solutions of YM equations then $ A_{\a} $ also
run over this set. The proposition is proved.

On one's way we have proved that Lagrangian $L$ is real and so
it can be written in the form

$$ L_{YM} = {1 \over 4e^2} \sqrt {-g}  ( \R A B {\a} {\b} R^{B
\ \a \b}_{\ A} + \hbox {c.c.} ) \eqno(36) $$

\ni where it is assumed that a part of functions $ \GG {\a} A B
$ are expressed through functions $ h_{AA'}, \rho $ from
identities (26) - (36).

Lagrangian $L$ with auxiliary condition of positivity of
Hermitian matrix $h_{AA'} $ give formulation of $SU(N)$ YM
theory in terms of $SU(N)$ gauge invariant variables $ \GG {\a}
A B , \ h_{AA'}, \ \rho $. But now we have new gauge symmetry
$GL(N,C)$. However, this gauge freedom can be easily eliminated
by imposing of covariant GU gauge. This will be demonstrated
in details for the case $N=2$ in the next section.

\vskip 23 pt

\noindent {\bf 3. $SU(2)$ Yang - Mills theory in gauge invariant
variables.}

\noindent {\bf 3.1.} {\boldit $GL(2,C)$ gauge fixing.}

\noindent  In this section we will consider the case $N=2$. Let
us impose the GU gauge

$$ \rho =1 \eqno(37) $$

Let

$$ (\e _{AB} ) = \left( \matrix {0 & 1 \cr -1 & 0 } \right)
\eqno(38) $$

\ni Below we will lower and rise capital Latin indexes $ A,B,...
=1,2 $ by metric spinor $  \e _{AB} $. For instance,

$$ \G _{\a AB}  = \e _{AC} \GG {\a} C A \eqno(39) $$

\ni It follows from (27) and (37), that $ \GG {\a} A A = 0. $ So

$$ \G _{\a AB} = \G _{\a BA} \eqno(40) $$

\ni and, consequently,

$$ R_{AB \a \b} = R_{BA \a \b } \eqno(41) $$

The Lagrangian (36) with $ \GG {\a} A B $ satisfying  (40) is
the Lagrangian of $ SL(2,C) $ gauge theory. But algebra
$sl(2,C)$ is isomorphic to algebra $ so(1,3) $. So our theory
can be reformulated in terms of $SO(1,3)$ gauge covariant
objects.

Due to isomorpfism $ sl(2,C) \approx so(1,3) $ $ \ $ any spinor
$ \f ^{AA'BB'...} $ corresponds to some $ SO(1,3) $ vector $ \f
^{ab...}. $ The correspondence is established by means of Infeld
- van  der  Vaerden symbols $ g^a_{\ AA'}(x) $:

$$ \f ^{ab...} = g^a_{\ AA'} g^a_{\ BB'} ... \f ^{AA'BB'...}
\eqno(42) $$

\ni Below, for short, we will write

$$ \f ^{ab...} = \f ^{AA'BB'...} \eqno(43) $$

\ni instead (42).

Quantities $ g^a_{\ AA'} $ can be choosen in different ways. In
particular, one can put

$$ g^a_{ \ AA'} = \hbox {const} \eqno(44) $$
$$ g^a_{\ AA'} g^{bAA'} =
\eta ^{ab} \eqno(45) $$

\ni where $ \eta ^{ab} =  \eta _{ab} = \hbox {diag} (+---) $
Quantities $ \eta ^{ab} $ and $  \eta _{ab} $ will be used for
lowering and raising indexes $ a,b,c,...=0,1,2,3.$

We would remind [7] that if

$$ \f ^{ab} = \f ^{AB} \e ^{A'B'} + c.c. \eqno(46) $$

\ni then $ \f ^{ab} $ is real antisymmetric tensor. So from
(40), (41) it follows that

$$ L = -{1 \over 8e^2} \sqrt {-g} R_{ab \a \b} R^{ab \a \b }
\eqno(47) $$

\ni where

$$ R^a_{\ b \a \b} = \d _{\a} \GG {\b} a b - \d _{\b} \GG {\a}
a b + \GG {\a} a c \GG {\b} c b - \GG {\b} a c   \GG {\a} c b
\eqno(48) $$

\ni and

$$ \G _{\a ab} = - \G _{\a ba} \eqno(49) $$

\ni Supplementary conditions (27), (37) can be written as

$$ D_{\a} h^a \equiv \d _{\a} h^a +  \GG {\a} a b h^b =0 \eqno(50)
$$

$$ h_a h^a =2 \eqno(51) $$

\ni for some real vector $h$.

In order to take onto account conditions (50), (51), we
introduce Lagrange multiplies $\la ^{\a}_a, \mu $ and consider
Lagrangian

$$ L_1 = L + \la ^{\a}_a D_{\a} h^a + \mu (h^2 - 2) \eqno(53) $$

\ni instead of Lagrangian $L$. In (53)  $\G _{\a ab}, h^a, \la
^{\a}_a, \mu $ are independent variables. But it is easy to
prove that on shell

$$\la ^{\a}_a =0, \ \ \ \mu =0  \eqno(54) $$

\ni So initial $SU(2)$ YM theory is equivalent to standard
$SO(1,3) $ theory with supplementary conditions (50),(51).
Therefore, due to results of the section 2, it is sufficient to
fix GU gauge for $SO(1,3)$ gauge theory in order to formulate
our theory in term of only gauge invariant variables. It is easy
to do in first order formalism.

In first order formalism Lagrangian (47) is equivalent to
Lagrangian

$$ L_2 = - {1 \over 4e} \sqrt {-g} \Phi _{ab \a \b } R^{ab \a
\b} + {1 \over 8} \sqrt {-g} \Phi _{ab \a \b } \Phi ^{ab \a \b }
\eqno(55) $$

\ni where variables $ \Phi ^{ab \a \b } =- \Phi ^{ba \a \b } =
 \Phi ^{ba \b \a} \hbox {and} \ \G _{\a ab} $ are considered as
independent ones. Let $ { \ }^+ \Phi ^+ _{ab \a \b} $ be
selfdual part of $  \Phi _{ab \a \b} $ with respect to both
pairs of indexes $ab$ and $\a \b $. We would remind that due to
isomorphism $so(1,3) \approx o(3,C) $ any selfdual antisymmetric
$SO(1,3)$ second rank tensor $\f ^{ab} $  can be described as
$O(3,C)$ vector $\f ^r = \f ^{0r} + {i \over 2} \e ^{rst} \f
^{st}, \ \ r,s,t =1,2,3. $ (In  Cartesian coordinate system in
Minkowsky space.) So tensor $ { \ }^+ \Phi ^+ _{ab \a \b} $ can
be considered as second rank $O(3,C)$ tensor $\Phi _{rs} $, and
we can impose desired GU gauge as follows:

$$   \Phi _{rs} = \Phi _{sr} \eqno(56) $$

This gauge is covariant analog Anishetty's gauge (21). It is
easy to prove that this gauge is regular on the set  $ \M _0 =
\{ \Phi _{sr} | \hbox {det} (\Phi _{sr}) \not \equiv 0 \} $.
By this note we finish our description of  $SU(2)$ YM theory in
gauge invariant variables.

\noindent {\bf 3.2} {\boldit Analogy with $R^2$-gravity.}

\noindent In the previous subsection we used Infeld - van der
Vaerden symbols $g^a_{\ AA'}$ defined by formulae (44), (45). In
this subsection we will use another choice of $g^a_{\ AA'}$ .
Namely, we will define $g^a_{\ AA'}$ so that

$$ g^a_{\ AA'} (x) g^{bAA'} (x) = g^{ab} (x)  \eqno(57) $$

\ni where $  g^{ab} (x)  $ is space - time metric tensor. This
choice of Infeld - van der Vaerden symbols entangles gauge
and space - time indexes and we will consider them on the equal
footing.

If (57) is hold, then

$$ D_{\a} g_{\b \g} \equiv \d _{\a} g_{\b \g} - \GG {\a} {\del} {\b}
g_{\del \g} - \GG {\a} {\del} {\g} g_{\b \del } =0 \eqno(58) $$

\ni So the connection $ \GG {\a} {\del} {\g} $ is metric. Let

$$ \GG {\a} {\b} {\g} = \left\{ { \b \atop \a \ \ \g } \right\}
+ Q^{\ \b}_{\a \ \g} \eqno(59) $$

\ni  where $ \left\{ { \b \atop \a \ \ \g } \right\} $ are usual
are usual Christoffel symbols. Substituting (59) in (47), one
obtains:

$$ L = - {1 \over 8e^2} \sqrt {-g} r_{\a \b \g \del } r^{\a \b
\g \del} + ...  \eqno(60) $$

\ni where $ r_{\a \b \g \del}$ is usual
Riemannien tensor defined by metric $g_{\a \b}$, and dots mean
terms proportional to nonzero degrees of $ Q^{\ \b}_{\a \ \g} $.

We see that $SU(2)$ YM fields induce $R^2$-gravity term

$$  - {1 \over 8e^2} \sqrt {-g} r_{\a \b \g \del } r^{\a \b
\g \del}   \eqno(61) $$

\ni in Lagrangian. In elementary particle physics terms of the
type (61) in Lagrangian were considered in recent papers [8],[9]
in connection with ideas of strong gravity [10]. Terms of the
type (61) are also considered in quantum gravity in connection
with the problems of renormalization [11], [12]. We see that
terms of such type can be obtained from standard $SU(2)$ YM
theory without any supplementary assumptions.

\vskip 23 pt

\noindent {\bf 5.Some additional notes.}

\noindent {\bf 5.1.} {\boldit The case $ N \geq 3 $ .}

\noindent If $ N \geq 3 $ then one can build the theory in two steps.
First, one can reduce $GL(N,C)$ gauge freedom to $SL(2,C)$ one
by imposing suitable GU conditions on $ h_{AA'}, \rho,$ and $
R^A_{\ B \a \b} $ (in first order formalism.) At the second steps
one can use ideas of section 4. Detailed investigation will be
presented in forthcoming paper.

\noindent {\bf 5.2.} {\boldit Matter fields.}

\noindent Introducing of matter fields is not a hard problem. If $
\Psi
^{rs...}_{MN...} $ are matter fields with colour indexes $
r,s,... $ and space - time spinor indexes $ M,N,...$ then one
can use decomposition

$$ \Psi ^{rs...}_{MN...} =\Psi ^{AB...}_{MN...}\E r A \E s B \
... \eqno(62) $$

\ni in order to introduce $SU(N)$ gauge invariant fields
$ \Psi ^{AB...}_{MN...} $ instead of initial fields $ \Psi
^{rs...}_{MN...} $ and express Lagrangian in terms of only $
SU(N)$ gauge invariant quantities.

\noindent {\bf 5.3.} {\boldit Three dimensional case.}

\noindent Results obtained in our previous paper [1] can be also
obtained
in framework of the scheme presented in the sections 2,3, but it
is necessary from the beginning to use first order formulation
and adjoined representation of $SU(2)$ instead of fundamental
one. Then, in notations of [1], analog of (62) can be written as
follows:

$$ { \ }^{*} F^a_i = \Phi ^A_{\ i} \E a A  \eqno(63) $$

\ni ($a=1,2,3, \ \ i,A=0,1,3.$) Results of [1] are reproduced
after imposing GU gauge

$$ \Phi ^A_{\ i} = \del ^A_{\ i}  \eqno(64) $$

\noindent {\bf 5.4.} {\boldit Eucleadian case. }

\noindent Formulation of Eucleadian $SU(2)$ in terms of only gauge
invariant variables can be obtained in the way analogous to one
presented in the previous subsection, but instead of $   { \
}^{*} F^a_i $ one must use (anti) selfdual part $ { \ }^{ \pm }
F^a_i = F^a_{0i} \pm \oh \e _{ijk} F^a_{jk} $ of field strength
tensor $ F^a_{\a \b} $ .

\vskip 23pt

\noindent {\bf 6. Conclusion. }

\noindent In previous sections we gave formulation of $SU(N)$  YM
theory
in terms of only gauge invariant variables. At least, in the
case $N=2$ this formulation appears to be very similar to
$R^2$-gravity and so it would be very interesting to develop our
investigation in connection with ideas of strong gravity [8] -
[10], microuniverses [13] and renormalization problems in
quantum gravity [11], [12].

Another interesting approach can be connected with the fact that
spin and isospin indexes are entangled in our final formulation
of YM theory and so they can be considered on the equal footing.
So isospin transforms in spin in our approach. In particular, for
solution of coupled  Einstein - Yang - Mills equations with
$Q^{\ \b}_{\a \ \g} =0 $ (see (59)) YM induced connection $\GG
{\a} {\b} {\g} $ coincides with usual metric connection $
\left\{ {\ \b \ \atop \a \ \g } \right\} $ and so spin and
isospin indexes are entirely "equalized in rights". In
paper [16] it was shown that such solutions exist but for only
certain values of coupling constant $e$.

The truck with expansion of gauge group can be repeated. In
particular, it is possible to prove that $SO(1,3)$ gauge theory
(that appears in our approach)  is equivalent to certain
$GL(4,R)$  gauge theory. This fact, may be, points out the
connection with the approach to particle classification based on
the group $GL(4,R)$ [14].

At quantum level our approach  automatically solves the problem
of colour confinement because the theory is formulated in terms
of only colourless quantities. Another attractive feature of our
formalism is the possibility to impose covariant boundary
conditions

$$ \Phi _{rs} \to \la \del _{rs}, \ \ \la =  \hbox {const}
\eqno(65) $$

\ni (see 56)) in functional integral, that, may be, can be used
for description of the gluon condensate.

\vskip 23 pt

\noindent {\bf Acknowledgements}

\noindent Author is indebted to {$ \ \ $ D.V.Gal'tsov, $ \ \ $
C.Grosse-Knetter, $ \ \ $ D.Z.Freedman, {$ \ \ $ M.B.Halpern, $
\ \ $ Y.Ne'eman, $ \ \ $ B.V.Medvedev, $ \  \ $A.A.Slavnov, $ \
\ $ N.A.Sveshnikov, $ \ \ $I.P.Volobuev , and Yu.M.Zinoviev for
interesting discussions and comments.

\vskip 23 pt

\noindent {\bf References}

{\ninerm
 \baselineskip =11 pt

\i {1} F.A.Lunev, {\nineit Phys. Lett.}, {\ninebf B295}, 99
(1992)

\i {2} F.A.Lunev, {\nineit Theor. Math. Phys.}, {\ninebf 94}, 66
(1993)

\i {3} L.V.Ovsyannikov, {\nineit Group Analyses of Differential
Equations}, Academic Press, 1982

\i {4} S.Weinberg, {\nineit Phys. Rev. Lett.}, {\ninebf 19}, 1264
(1967); A.  Salam, in: {\nineit Elementary particle theory},
Almquist Furlag AB, 1968

\i {5} V.A.Matveev, A.N.Tavkhelidze, M.E.Shaposhnikov,
{\nineit Theor.  Math. Phys.}, {\ninebf 59}, 323 (1984)

\i { 6 } R.Anishetty, {\nineit Phys. Rev.}, {\ninebf D44}, 1895 (1991)

\i { 7 } R.Penrose, R.Rindler, {\nineit Spinors and space - time,
vol.  1}, Cambridge University Press, 1984

\i { 8 } A.Salam, C.Sivaram, {\nineit Mod. Phys. Lett.}, {\ninebf A8},
321 (1993)

\i { 9 } Y.Ne'eman, Dj.Sijacki, {\nineit Phys. Lett.} {\ninebf B247},
571 (1990); {\ninebf 276}, 173 (1992)

\i { 10 } A.Salam, {\nineit ICTP preprint IC/71/3}, (1971);
C.J.Isham, A.Salam, J.Strathdee, {\nineit Phys. Rev.}, {\ninebf D8},
2600 (1973); C.Sivaram, K.P.Singha, {\nineit Phys.Rep.}, {\ninebf 51},
111 (1979)

\i { 11 } B.S. DeWitt, {\nineit Dynamical Theory of Groups and
Fields}, Gordon and Breach, 1965

\i { 12 } K.S.Stelle, {\nineit Phys. Rev.}, {\ninebf D16}, 963 (1977)

\i { 13 } A.Salam, J.Strathdee, {\nineit Phys. Rev.}, {\ninebf D16},
2668 (1977)

\i { 14 } Y.Ne'eman, Dj.Sijacki, {\nineit Phys. Lett.}, {\ninebf
B157},
267 (1985); {\nineit Phys. Rev.}, {\ninebf D37}, 3267 (1988)

\i { 15 } F.A.Lunev, {\nineit J. Nucl. Phys.}, {\ninebf 56}, 238
(1993)

\i { 16 } F.A.Lunev, {\nineit Phys. Lett.}, {\ninebf B314}, 21 (1993)

\i { 17 } K.Johnson, in: {\nineit $ \ \ $QCD - 20 $ \ $Years
$ \ $ Later}, $ \ $Aahen, $ \ $ June 1992; $ \ $ D.Z.Freedman,
P.E.Haagensen, K.Johnson, J.I.Lattore, {\it Cern preprint
TH.7010/93}

\i { 18 } C.Grosse-Knetter, R.Kogerler, {\nineit Phys. Rev. }, {\bf
D48 }, 2865 (1993)

\i { 19 } F.G.Scholz, G.B.Tupper, {\nineit Phys. Rev.}, {\ninebf D48},
1792 (1993)

\i { 20 } J.Goldstone, R.Jackiw, {\nineit Phys. Lett.}, {\ninebf B74},
81 (1978)

\i { 21 } M.B.Halpern, {\nineit Phys. Rev.}, {\ninebf D16}, 1798, 3515
(1977); {\nineit Nucl. Phys.}, {\ninebf B139}, 477 (1978); {\nineit
Phys. Rev.}, {\ninebf D19}, 517 (1979)

\bye